\documentclass{emulateapj}

\usepackage{natbib}
\usepackage{amsmath}

\usepackage{txfonts}
\usepackage{float}
\usepackage{hyperref}
\usepackage[english]{babel}

\slugcomment{ApJ Letters in press}

\shorttitle{A unified empirical model for infrared galaxy counts based on observed physical evolution of distant galaxies}
\shortauthors{B\'ethermin et al.}

\begin{document}

\title{A unified empirical model for infrared galaxy counts based on the observed physical evolution of distant galaxies}




\author{Matthieu B\'ethermin\altaffilmark{1}, Emanuele Daddi\altaffilmark{1}, Georgios Magdis\altaffilmark{2}, Mark T. Sargent\altaffilmark{1}, Yashar Hezaveh\altaffilmark{3}, David Elbaz\altaffilmark{1}, Damien Le Borgne\altaffilmark{4,5}, James Mullaney\altaffilmark{1}, Maurilio Pannella\altaffilmark{1}, V\'eronique Buat\altaffilmark{6}, Vassilis Charmandaris\altaffilmark{7}, Guilaine Lagache\altaffilmark{8}, and Douglas Scott\altaffilmark{9}}

\altaffiltext{1}{Laboratoire AIM-Paris-Saclay, CEA/DSM/Irfu - CNRS - Universit\'e Paris Diderot, CEA-Saclay, Orme des Merisiers, F-91191 Gif-sur-Yvette, France, Email: \url{matthieu.bethermin@cea.fr}}
\altaffiltext{2}{Department of Physics, University of Oxford, Keble Road, Oxford OX1 3RH, UK}
\altaffiltext{3}{Department of Physics, McGill University, 3600 Rue University, Montreal, Quebec H3A 2T8, Canada}
\altaffiltext{4}{UPMC Univ. Paris 06, UMR7095, Institut d'Astrophysique de Paris, 75014 Paris, France}
\altaffiltext{5}{CNRS, UMR7095, Institut d'Astrophysique de Paris, 75014 Paris, France}
\altaffiltext{6}{Laboratoire d'Astrophysique de Marseille, OAMP, Universit\'e Aix-Marseille, CNRS, 38 rue Fr\'ed\'eric Joliot-Curie, 13388 Marseille
Cedex 13, France}
\altaffiltext{7}{Department of Physics \& Institute of Theoretical and Computation Physics, University of Crete, 71003 Heraklion, Greece}
\altaffiltext{8}{Institut d'Astrophysique Spatiale (IAS), b\^atiment 121, Universit\'e Paris-Sud 11 and CNRS (UMR 8617), 91405 Orsay, France}
\altaffiltext{9}{Department of Physics \& Astronomy, University of British Columbia, 6224 Agricultural Road, Vancouver, BC V6T 1Z1, Canada}

\begin{abstract}
We reproduce the mid-infrared to radio galaxy counts with a new empirical model based on our current understanding of the evolution of main-sequence (MS) and starburst (SB) galaxies. We rely on a simple Spectral Energy Distribution (SED) library based on \textit{Herschel} observations: a single SED for the MS and another one for SB, getting warmer with redshift. Our model is able to reproduce recent measurements of galaxy counts performed with \textit{Herschel}, including counts per redshift slice. This agreement demonstrates the power of our 2 Star-Formation Modes (2SFM) decomposition for describing the statistical properties of infrared sources and their evolution with cosmic time. We discuss the relative contribution of MS and SB galaxies to the number counts at various wavelengths and flux densities. We also show that MS galaxies are responsible for a bump in the 1.4~GHz radio counts around 50~$\mu$Jy. Material of the model (predictions, SED library, mock catalogs...) is available online\footnote{at http://irfu-i.cea.fr/Pisp/matthieu.bethermin/}.

\end{abstract}


\section{Introduction}

Recent observational studies have shown that two distinct star-forming (SF) mechanisms are required to describe the SF galaxy population. The so-called SF main sequence (MS) is composed of secularly-evolving galaxies that display a tight correlation between stellar mass ($M_\star$) and star formation rate (SFR) at a given redshift \citep[e.g.][]{Elbaz2007,Noeske2007,Daddi2007}. This population accounts for $\sim$85\% of the star formation rate density (SFRD) in the Universe \citep{Rodighiero2011,Sargent2012} at z$<$2. The rest of the star-formation budget is provided by starbursts (SB), i.e. galaxies with very high specific star formation rates (sSFR=SFR/$M_\star$), probably induced by recent mergers \citep[e.g.][]{Elbaz2011,Rodighiero2011}. Recently, \citet[][S12 hereafter]{Sargent2012} showed that infrared (IR) luminosity functions (LF) can be reproduced by jointly considering the mass function of SF galaxies (SFMF), the evolution of the sSFR of MS galaxies, and its distribution at fixed M$_\star$, with a separate contribution from MS and SB galaxies.\\

Wavelength-dependent galaxy number counts are an additional, important constraint for evolutionary models of infrared galaxies. While purely semi-analytical models \citep[e.g.][]{Lacey2010,Somerville2011} struggle to reproduce infrared (IR) number counts, phenomenological or hybrid models \citep[e.g.][]{Bethermin2011,Gruppioni2011,Rahmati2011,Lapi2011} fare better but are in general descriptive and use an evolution of the luminosity function which is not motivated by physical principles. However, these recent models which reproduce the total counts passably, are excluded at $>$3\,$\sigma$ by the recent \textit{Herschel} measurements of counts per redshift slice \citep{Berta2011,Bethermin2012}. This shows how important redshift-dependent constraints are to accurately model the evolution of galaxies, and motivates the development of a new generation of models.\\



We present a new model of IR galaxy counts which builds on the 2-Star-Formation-Mode framework (2SFM) S12 introduced. This fiducial model is intuitive and based on our current observational knowledge of the evolution of MS and SB galaxies. All model parameters are constrained by external datasets and require no additional fine-tuning. We assume a Salpeter initial mass function and a \textit{WMAP}-7 cosmology.\\



\section{Main ingredients}

\label{sect:ingredients}

Our model is based on four main ingredients, which are sufficient to reach a good agreement with IR source counts (see Sect.~\ref{sect:results} and grey line Fig.~\ref{fig:counts}):
\begin{itemize}
\item evolution of the MS with redshift,
\item decomposition of the sSFR distribution at fixed $M_\star$ into MS and SB modes,
\item evolution of the SFMF with redshift,
\item Spectral Energy Distribution (SED) libraries for MS and SB galaxies.
\end{itemize}
Additional ingredients, which are of lesser importance, are presented in Sect.~\ref{sect:refinements}.

\begin{table*}
\centering
\caption{\label{tab:par} Summary of parameters of our fiducial model.}
\begin{tabular}{llrl}
\hline
\hline
Parameter & Description & Value & Reference \\
\hline
\multicolumn{4}{c}{Distribution of sSFR}\\
\hline
$B_{SB}$ & Boost of specific star formation rate in SB (in dex) & 0.6 & Sargent et al. (2012, hereafter S12)\\
$\sigma_{MS}$ & Width of the MS log-normal distribution (in dex) & 0.15 & S12 value minus 0.05\,dex for artificial scatter \citep{Salmi2012}\\
$\sigma_{SB}$ & Width of the SB log-normal distribution (in dex) & 0.20 & S12 value minus 0.05\,dex for artificial scatter \citep{Salmi2012}\\
\hline
\multicolumn{4}{c}{Evolution of the main MS}\\
\hline
sSFR$_{MS,0}$ & sSFR on the MS at $z=0$ and $M_\star=10^{11}\,M_\odot$ (in log(yr$^{-1}$)) & $-10.2$ & S12 based on a fit of literature data\\
$\beta_{MS}$ & Slope of the sSFR-M$_\star$ relation at a given redshift & --0.2 & \citet{Rodighiero2011} and S12\\
$\gamma_{MS}$ & Evolution of the normalization of the MS with redshift & 3 & compilation of measurements of sSFR (see Fig.~\ref{fig:evo})\\
$z_{evo}$ & Redshift where the MS normalization stops to evolve & 2.5 & compilation of measurements of sSFR (see Fig.~\ref{fig:evo})\\
\hline
\multicolumn{4}{c}{Evolution of the fraction of SB}\\
\hline
$r_{SB,0}$ & Relative amplitude of SB log-normal distribution compared to MS & 0.012 & S12 and \citet{Hopkins2010}  (see Fig.~\ref{fig:evo})\\
$\gamma_{SB}$ & Evolution of starburst fraction with redshift & 1 &  S12 and \citet{Hopkins2010} (see Fig.~\ref{fig:evo})\\
$z_{SB}$ & Redshift where the starburst fraction stops to evolve & 1 & S12 and \citet{Hopkins2010} (see Fig.~\ref{fig:evo})\\
\hline
\multicolumn{4}{c}{Star-forming mass function (SFMF) and its evolution}\\
\hline
$M_b$ & Stellar mass at the break of the SFMF (in log(M$_\odot$)) & 11.20 & \citet{Ilbert2010} and \citet{Peng2011}\\
$\alpha$ & Faint-end slope of the SFMF & 1.3 & \citet{Ilbert2010} and \citet{Peng2011}\\
$\phi_b(z<1)$ & Number density at the break of the SFMF at $z<1$ (in log(Mpc$^{-3}$)) & --3.02 & \citet{Sargent2012} (see Fig.~\ref{fig:evo})\\ 
$\gamma_{SFMF}$ & Evolution of the density of SFMF at $z>1$ & 0.4 & Extended from \citet{Sargent2012} (see Fig.~\ref{fig:evo})\\
\hline
\multicolumn{4}{c}{Evolution of SEDs}\\
\hline
$\langle U  \rangle_{MS,0}$ & Mean radiation field in local MS galaxies & 4 & \citet{Magdis2012} and Fig.\,\ref{fig:sed}\\
$\gamma_{U,MS}$ & Evolution of the radiation field in MS with redshift  & 1.3 & \citet{Magdis2012}  and Fig.\,\ref{fig:sed}\\
$z_{\langle U  \rangle,MS}$ & Redshift where $\langle U  \rangle$ in MS flattens & 2 & \citet{Magdis2012}  and Fig.\,\ref{fig:sed}\\
$\langle U  \rangle_{SB,0}$ & Mean radiation field in local SB galaxies & 35 & \citet{Magdis2012}  and Fig.\,\ref{fig:sed}\\
$\gamma_{U,SB}$ & Evolution of the radiation field in SB with redshift  & 0.4 & \citet{Magdis2012}  and Fig.\,\ref{fig:sed}\\
$z_{\langle U  \rangle,SB}$ & Redshift where $\langle U  \rangle$ in SB flattens & 3.1 & \citet{Magdis2012}  and Fig.\,\ref{fig:sed}\\
\hline
\multicolumn{4}{c}{Contribution of AGNs}\\
\hline
$A_{AGN}$ & Normalization of the probability distribution of AGN contribution & 0.0025 & \citet{Aird2012}\\ 
$\beta_{AGN}$ & Slope of the probability distribution of AGN contribution & --0.7 & See Sect.~\ref{sect:refinements}\\
\hline
\end{tabular}
\end{table*}

\subsection{SFR distribution}

\begin{figure}
\centering
\includegraphics[width=8.5cm]{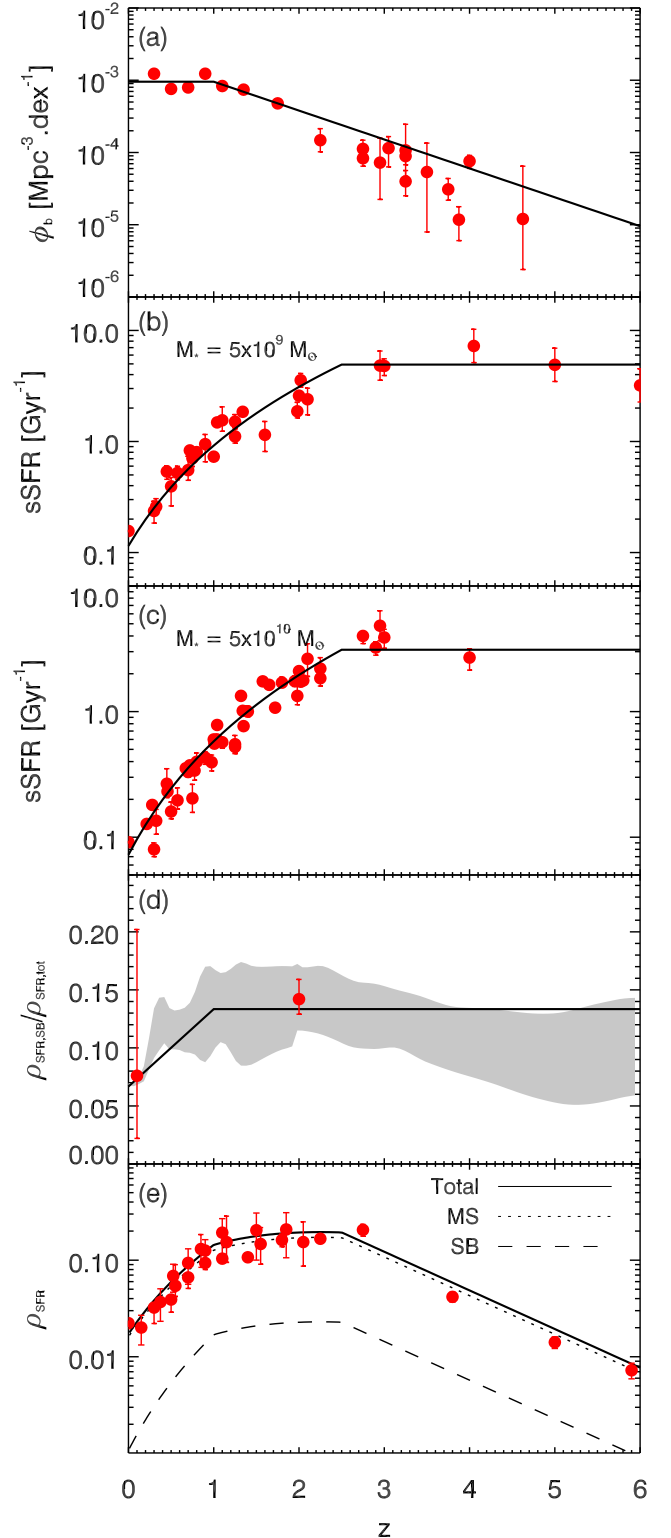}
\caption{\label{fig:evo} Redshift evolution of selected model parameters and derived quantities; model conventions are represented using a solid line; (a) -- density at the break of the mass function; (b) -- sSFR at $M_{\star}=5\times10^9\,M_\odot$. (c) -- sSFR at $M_{\star}=5\times10^{10}\,M_\odot$ (data in a, b, and c from a compilation of Sargent et al. in prep.); (d) -- contribution of SB to the total star formation rate density (data from S12; grey region from \citealt{Hopkins2010}). (e) -- SFRD. The MS (SB) contribution is represented by a dotted (dashed) line (data: \citet{Bouwens2007}, \citealt{Rodighiero2011}, \citealt{Magnelli2011} and \citealt{Karim2011}).}
\end{figure}

A key ingredient of the S12 approach is the probability distribution of sSFR at fixed M$_\star$ for SF galaxies based on observations presented in \citet{Rodighiero2011}. It is parametrized as a double log-normal decomposition of MS and SB:
\begin{equation}
\begin{split}
p(\textrm{log(sSFR)}) \propto \textrm{exp} \left ( -\frac{ \left (\textrm{log(sSFR)}-  \textrm{log(sSFR}_{MS}) \right )^2}{2 \sigma_{MS}^2} \right ) \\
+ r_{SB} \times \textrm{exp} \left (-\frac{ \left ( \textrm{log(sSFR)}- \textrm{log(sSFR}_{MS}) - B_{SB}) \right )^2}{2 \sigma_{SB}^2} \right ),
\end{split}
\end{equation}
where $\sigma_{MS}$ and $\sigma_{SB}$ are the dispersion in the sSFR of the MS and the SB populations. $B_{SB}$ is the average sSFR-boost for SB galaxies. We assume that these three parameters do not evolve with $M_\star$ and redshift, as suggested by S12 who reproduce the z$\sim$0 IR LF under these assumptions and with the distribution calibrated at z$\sim$2 (see Table~\ref{tab:par} for parameter values adopted). sSFR$_{MS}$ varies with $M_\star$ and redshift according to
\begin{equation}
\begin{split}
\textrm{sSFR}_{MS} (z,M_\star) = & \textrm{sSFR}_{MS,0} \times \left ( \frac{M_\star}{10^{11}\,M_\odot} \right)^{\beta_{MS}} \\
	& \times  \left ( 1+ \textrm{min} \left (z, z_{evo} \right ) \right )^{\gamma_{MS}},
\end{split}
\end{equation}
where $sSFR_{MS,0}$ is the sSFR at z=0 for $M_\star = 10^{11}\,M_\odot$ and $\beta_{MS}$ parametrizes the dependance of sSFR on $M_\star$. $\gamma_{MS}$ describes the evolution of the normalization of the MS out to redshift $z_{evo}=2.5$ where this evolution flattens according to observations (e.g. \citealt{Gonzalez2010}). The values of these parameters, chosen based on measurements summarized in Fig.\,\ref{fig:evo}bc, are listed in Table\,\ref{tab:par}. S12 also present evidence for a weak redshift evolution of r$_{\rm SB}$, the relative amplitude of SB sSFR log-normal distribution compared to MS one (or, equivalently, of the relative SB-contribution to the SFRD, see Fig.\,\ref{fig:evo}d), in agreement with the model of \citet{Hopkins2010}. Here we define the redshift evolution of $r_{SB}$ as:
\begin{equation}
r_{SB}(z) = r_{SB,0} \times \left ( 1+ \textrm{min} \left (z, z_{SB} \right ) \right )^{\gamma_{SB}}, \, \textrm{where} \, z_{SB} = 1,
\end{equation}
in order to broadly reproduce the trends suggested by these two studies (see Fig.~\ref{fig:evo}d). The impact of this evolving $r_{SB}$ is negligible, barring a $\sim$20\% decrease of 70\,$\mu$m counts compared to a constant $r_{SB}$.\\

Another important ingredient of our model is the evolution of the SFMF. Observations are well-described by a Schechter function
\begin{equation}
\phi = \frac{dN}{d \textrm{log}(M_\star)} = \phi_b (z) \times \left ( \frac{M_\star}{M_b} \right )^{- \alpha_{MF}} \times \textrm{exp} \left ( -  \frac{M_\star}{M_b} \right ) \times  \frac{M_\star}{M_b} \textrm{ln}(10)
\end{equation}
with a redshift-invariant characteristic mass $M_b$ and faint-end slope $\alpha_{MF}$, in keeping with \citet{Peng2011}. $\phi_b$, the characteristic density, is constant between z=0 and z=1 but decreases at z$>$1 as
\begin{equation}
\textrm{log}(\phi_b) = \textrm{log}(\phi_b)(z<1) + \gamma_{SFMF} (1-z).
\end{equation}
The fiducial values (chosen from Fig.~\ref{fig:evo}a) of the MF-related parameters are also listed in Table~\ref{tab:par}.\\

The star formation history implied by our evolutionary formalism is shown in Fig.\,\ref{fig:evo}e. The star formation rate density increases from z=0 to 1, flattens between $z=1$ and $z=z_{evo}=2.5$ and decreases with redshift at $z>z_{evo}$, matching the infrared measurements of \citet{Magnelli2011} and \citet{Rodighiero2011}, the radio measurements of \citet{Karim2011}, and the optical measurements of \citet{Bouwens2007} at high redshift. The SFMF is quite uncertain at $z>4$, but this has little impact on the counts. In our model, the SFRD is dominated by MS galaxies at all redshifts.\\

\subsection{SEDs}

\begin{figure}
\centering
\includegraphics[width=9cm]{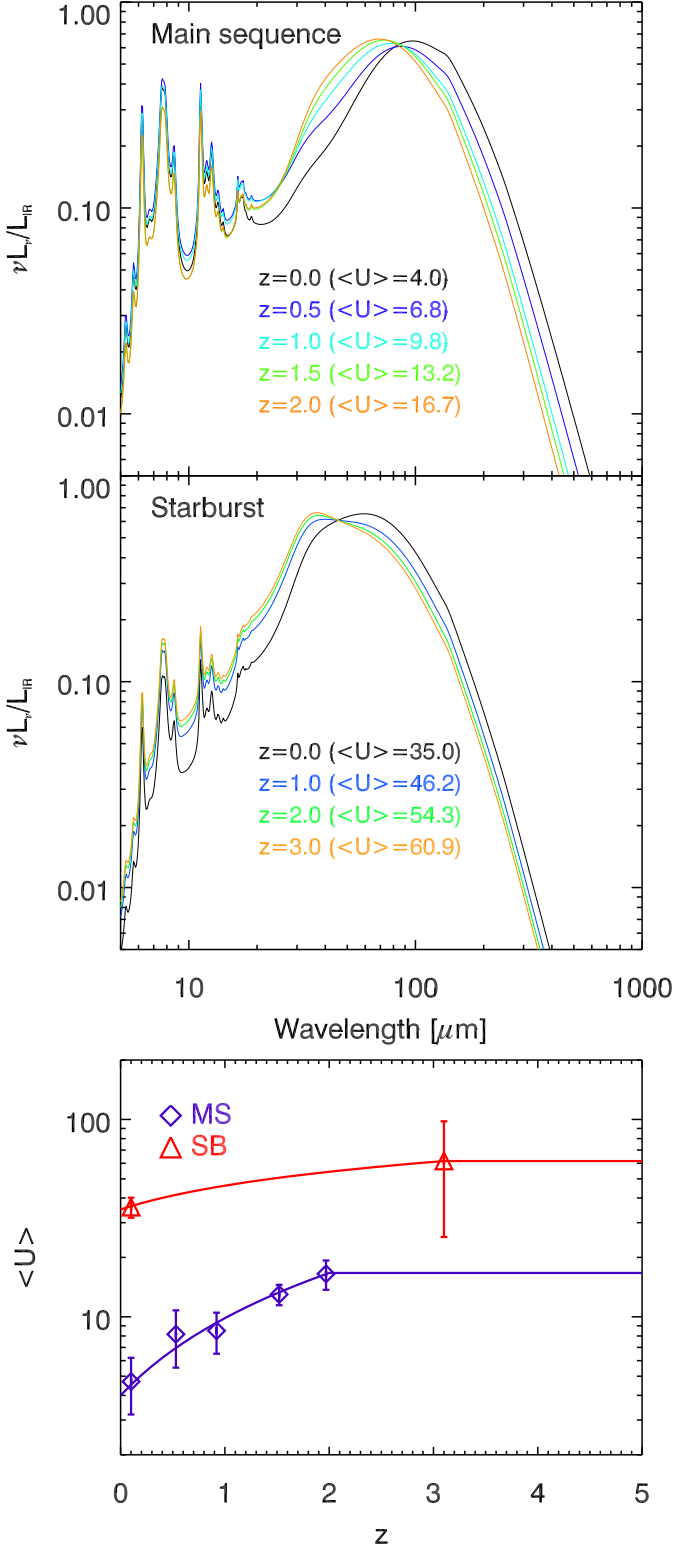}
\caption{\label{fig:sed} SEDs (normalized to have $L_{IR}=1$\,L$_\odot$) for MS (top) and SB (middle) used in our model. Evolution of $\langle U \rangle$ parameter with redshift in MS (blue) and SB (red) galaxies (data from M12).}
\end{figure}

We use a characteristic IR SED template for MS and SB based on fits of \citet{Draine2007} models to \textit{Herschel} observations of distant galaxies as presented in \citet[][hereafter M12]{Magdis2012}. While conceptually similar to \citet{Elbaz2011}, however, our templates evolve with redshift following the finding of M12 that the mean radiation field $\langle U \rangle$ (which correlates with dust temperature) is more intense at high redshift:
\begin{equation}
\langle U  \rangle = \langle U  \rangle_0 \times \left ( 1+\textrm{min}(z,z_{\langle U \rangle}) \right )^{\gamma_U}.
\end{equation}
Here $\langle U  \rangle_0$ is the mean radiation field in local MS galaxies, $\gamma_{U}$ a parameter determining its evolution with redshift, and $z_{\langle U  \rangle}$ the redshift where $\langle U  \rangle$ flattens. This evolution is different in MS and SB galaxies (see Fig.\,\ref{fig:sed} and Table\,\ref{tab:par}). This evolution is caused by the evolution of SF efficiency and metallicity with redshift (M12), and is required to reproduce source counts. For example, if we used the z=1 (z=0) MS template for all redshifts, we would overestimate the counts by about a factor of 2 (1) at 70\,$\mu$m and 2 (10) at 1.1\,mm. For reference, if we use the MS and SB templates of \citet{Elbaz2011}, we overpredict the mm counts by a factor of 10 at all fluxes and underpredict the 100\,$\mu$m counts by $\sim$30\%. To reproduce the 24\,$\mu$m counts, it is crucial to use distinct SB templates with less mid-IR emission than in MS galaxies. The SEDs of MS and SB galaxies used in our model are shown in Fig.~\ref{fig:sed}. We introduce a relative dispersion on $\langle U \rangle$ of 0.2\,dex for both MS and SB (M12), which has little impact on the counts ($<$10\,\%), except in the millimeter domain ($+20\%$). In this approach, the increasing mean dust temperature with infrared luminosity ($L_{IR}$) at a given redshift is caused by a higher fraction of SB galaxies at higher $L_{IR}$.\\

\section{Refinements}

\label{sect:refinements}

\subsection{Dust attenuation}

To reproduce IR number counts we have to link SFR and $L_{IR}$. For obscured SF galaxies, the bulk of the UV light emitted by young stars is absorbed by dust and re-emitted in the IR (SFR$_{IR}$/$L_{IR}$ = K = 1.7$\times10^{-10}$\,M$_\odot$yr$^{-1}$L$_\odot^{-1}$, \citealt{Kennicutt1998}). In less massive galaxies, the attenuation is smaller and a significant part of the SF can be detected in UV. The total star formation can then be divided into an uncorrected UV and an IR component ($SFR = SFR_{UV}+SFR_{IR}$). The mean ratio between these two components, r$_{1500}$, varies with M$_\star$. Here, we apply the relation of \citet{Pannella2009}:
\begin{equation}
r_{1500} = 2.5 \textrm{log} \left ( \frac{\textrm{SFR}_{IR}}{\textrm{SFR}_{UV}} \right ) = 4.07 \times \textrm{log}\left(\frac{M_\star}{M_\odot} \right) - 39.32,
\end{equation}
and assume redshift-invariance, as suggested by \citet{Sobral2012} and Pannella et al. (in prep.). The IR luminosity of the galaxies, $L_{IR}$ is thus given by:
\begin{equation}
L_{IR} = \frac{\textrm{SFR}_{IR}}{K} = \frac{\textrm{SFR}}{K} \times \frac{10^{0.4 \times r_{1500}}}{1+10^{0.4 \times r_{1500}}} = \frac{\textrm{SFR}}{K} \times f_{IR}^{SF}(M_\star),
\end{equation}
where $f_{IR}^{SF}(M_\star)=SFR_{IR}/SFR$ goes to 0 at low mass and 1 at high mass. This correction implies a flatter IR LF at the faint end as compared to the SFMF at the low-mass end and prevents an excess in the counts at faint flux densities. Although a small part of the IR emission is due to dust heated by old stars, especially at low-z, we consistently reproduce $z=0-2$ IR LF (S12).\\

\subsection{AGN contribution}

\begin{figure*}
\centering
\includegraphics[width=17cm]{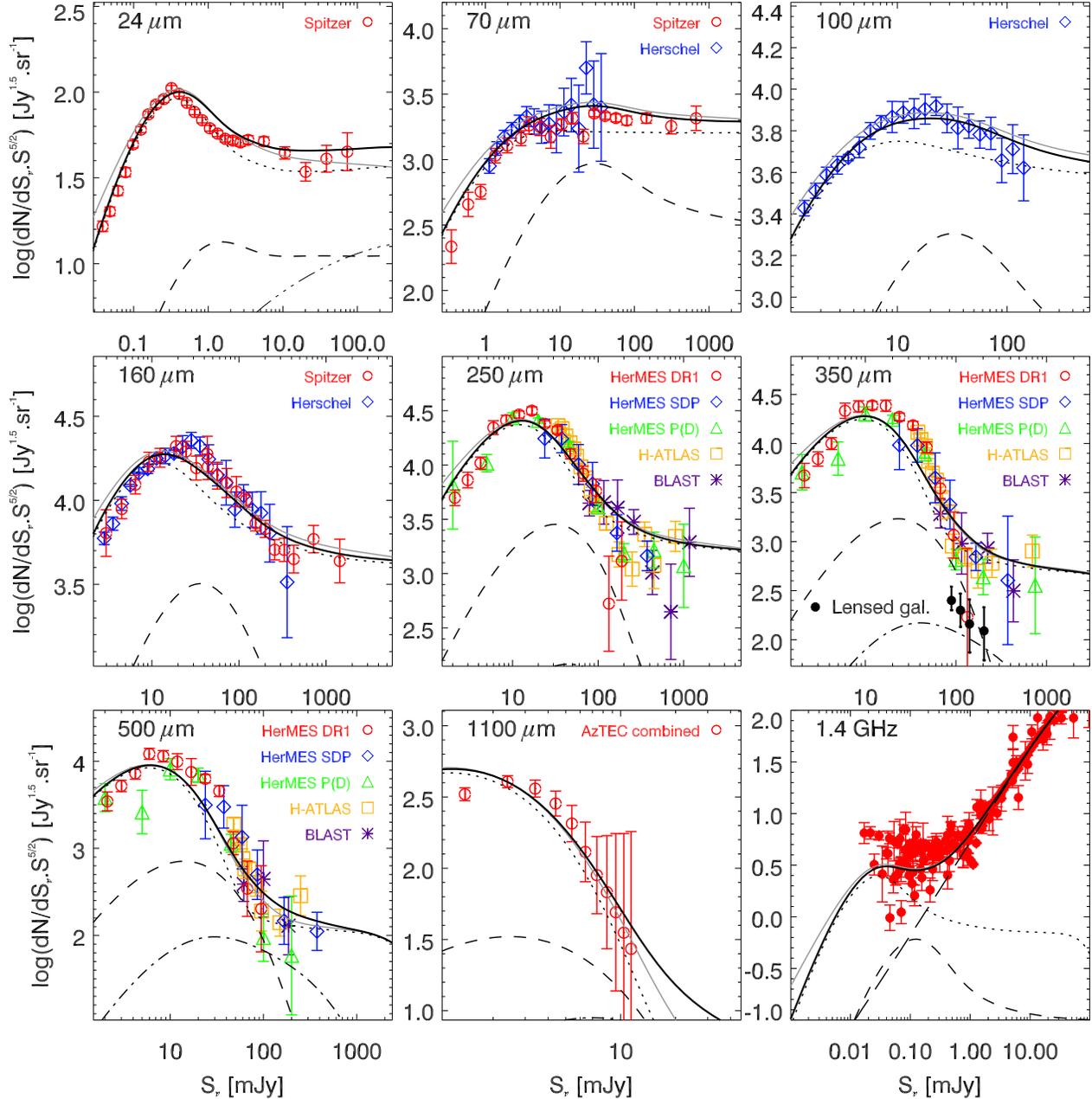}
\caption{\label{fig:counts} Number counts from 24\,$\mu$m to 1.4\,GHz. Solid line-- total counts predicted by the model; grey line -- counts predicted by the simplified model (without refinements discussed in Sect.~\ref{sect:refinements}); dotted line -- MS contribution; short-dashed line-- SB contribution;  dot-dashed line -- lensed sources; triple-dot-dashed line-- difference between counts with and without AGN contribution. At 1.4\,GHz, we also plot the model of AGN-driven radio sources of \citet{Massardi2010} (long-dashed line) and combine it with our model of SF galaxies. Data points -- \citet{Bethermin2010a} (red points at 24, 70, and 160\,$\mu$m), \citet{Berta2011} (blue points 70, 100, and 160\,$\mu$m), \citet{Bethermin2012} (red points at 250, 350 and 500\,$\mu$m), \citet{Oliver2010} (blue points at 250, 350 and 500\,$\mu$m), \citet{Glenn2010} (green points at 250, 350 and 500\,$\mu$m), \citet{Clements2010} (yellow points at 250, 350 and 500\,$\mu$m), \citet{Bethermin2010b} (purple points at 250, 350 and 500\,$\mu$m), \citet{Scott2012} (red points at 1.1\,mm), and \citet{Vernstrom2011} (compilation of 1.4\,GHz radio counts). Black dots -- contribution of lensed galaxies at 350\,$\mu$m measured by \citet{Gonzalez-Nuevo2012}.}
\end{figure*}

Active-Galactic-Nucleus (AGN) activity is potentially important when modeling mid-IR counts. We statistically associate an AGN contribution, represented by the average intrinsic SED template of \citet{Mullaney2011}, to each galaxy based on its $L_{IR}$. \citet{Aird2012} showed that the Eddington ratio $r_{\textrm{Edd}}$ (bolometric luminosity $L_{bol}^{AGN}$ over Eddington luminosity) of AGN at z$<$1 follows a power-law probability distribution function (PDF) with redshift-dependent normalization. Based on the results of \citet{Mullaney2012b} -- who report a coincident cosmological evolution of the averages of specific black hole (BH) growth ($\dot{M}_{BH}/M_{BH}$, where $M_{BH}$ is BH mass) and sSFR over 0.5$<$z$<$2.5, a fact that implies constant $M_{BH}$/$M_{\star}$ ratios -- , we can express the \citet{Aird2012} results in terms of a distribution of ratios of bolometric luminosities ($r_{AGN} = L_{IR}^{AGN}/L_{IR}^{SF}$) from AGN and SF with redshift-independent normalization:
\begin{equation}
p(r_{Edd}) = C(z) \times r_{Edd}^{\beta_{AGN}} \rightarrow p(r_{AGN}) = A_{AGN} \times r_{AGN}^{\beta_{AGN}},
\label{eq:agn}
\end{equation}
where we recall that
\begin{equation}
\frac{\dot{M}_{BH}}{M_{BH}} \propto \frac{L_{bol}^{AGN}}{M_\star} \propto \frac{L_{IR}^{AGN}}{(L_{IR}^{SF})^{1+\beta_{MS}}} \approx \frac{L_{IR}^{AGN}}{L_{IR}^{SF}}.
\end{equation}
The last step uses the $M_\star$-SFR correlation. ${\beta_{AGN}}=-0.7$ comes from \citet{Aird2012}. $A_{AGN}$ is based on the normalization of the \citet{Aird2012} relation and includes a scaling factor for the conversion between $r_{Edd}$ and $r_{AGN}$ PDFs. This scaling relation assumes a mean ratio between black hole and stellar mass of 0.0015 \citep{Mullaney2012b}, plus a mean ratio between $L_{IR}^{AGN}$ and $L_{bol}^{AGN}$ calibrated from \citet{Mullaney2011}, \citet{Lutz2004} and  \citet{Vasudevan2007}. In order to normalize this PDF, we place a cut at $\lambda_{Edd} = 1$ and choose a lower bound such that $\int p(r_{AGN}) dr_{AGN}=1$. We emphasize that Eq.\,\ref{eq:agn} implies a correlation between AGN and SF activity only in an average sense, while preserving a large dispersion for individual objects consistent with observations. Full details of our AGN-treatment will be presented in a future paper. The AGN contribution is significant ($>10\%$) only at 24\,$\mu$m above 3\,mJy (see Fig.\,\ref{fig:counts}) and negligible at longer wavelengths ($<2\%$).\\

\subsection{Magnification caused by strong lensing}

Having computed the IR LF, split into MS and SB contribution as in S12, we include the effect of the strong ($\mu>2$) lensing  \citep{Negrello2007, Negrello2010} on these two LFs:
\begin{equation}
\frac{d^2 N}{d\textrm{log}L_{IR} dV} \bigl |_{\textrm{lensed}}  = \int_{\mu = 2}^{\infty} \frac{dP(\mu,z)}{d \textrm{log} \mu} \frac{d^2 N}{d\textrm{log}L_{IR} dV} \bigl |_{\textrm{initial}} d \textrm{log} \mu,
\end{equation}
where $\mu$ is the magnification, $\frac{dP}{d \textrm{log} \mu}$ the magnification PDF in the \citet{Hezaveh2011} model, and $\frac{d^2 N}{d\textrm{log}L_{IR} dV}$ the luminosity function. These lensed sources contribute $\sim$20\% to (sub-)mm counts around 100~mJy. 


\begin{figure*}
\centering
\includegraphics[width=17cm]{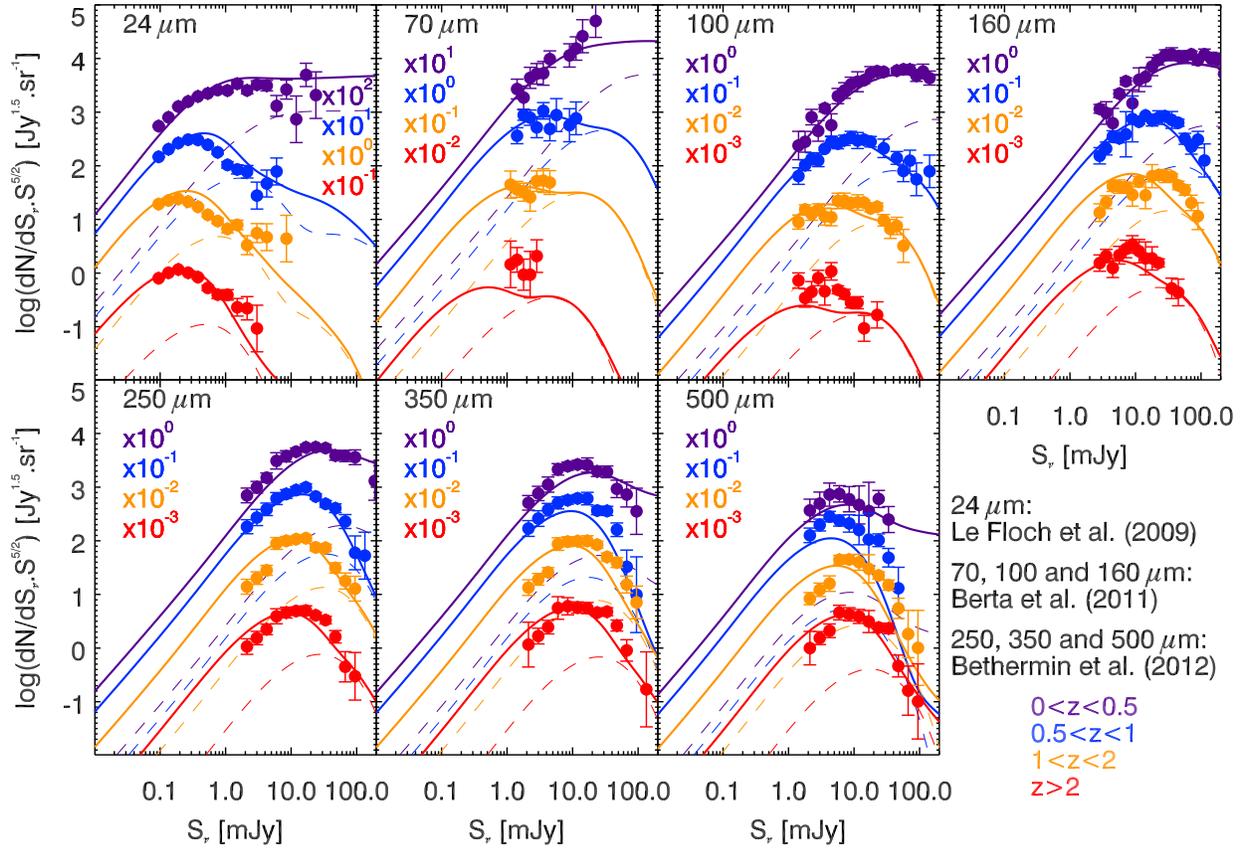}
\caption{\label{fig:counts_zslice} Normalized number counts per redshift slice, compared to model predictions. Data from \citet{Le_Floch2009} (24\,$\mu$m), \citet{Berta2011} (70, 100, 160\,$\mu$m), and \citet{Bethermin2012} (250, 350, 500\,$\mu$m). For clarity, a redshift-dependent vertical offset has been applied to model and data. Dashed line -- contribution of SB.}
\end{figure*}

\section{Results}

\label{sect:results}

Number counts are computed according to:
\begin{equation}
\begin{split}
\frac{d^2 N}{d S dz}(S,z,\lambda)  = \sum_{\textrm{type = \{MS, SB\}}} \int_{\langle U \rangle} \int_{r_{AGN}} d r_{AGN} \, d \langle U \rangle\\
\times \frac{d^2 N_{\textrm{type}}}{dL_{IR} dV} \left (z, L_{IR}\left (S, \textrm{type},\langle U \rangle, z \right ) \right )\\
\times \frac{1\,L_\odot}{S_{\textrm{norm}}^{\textrm{type}, \langle U \rangle}(z,\lambda) + r_{AGN} \times S_{\textrm{norm}}^{AGN} (z,\lambda)} \, \frac{dV}{dz}\,  p(r_{AGN}) \, p(\langle U \rangle | z, \textrm{type}) \\ 
\end{split}
\end{equation}
Here $S_{\textrm{norm}}^{\textrm{type}, \langle U \rangle}$ is the flux of a $L_{IR}=1\,L_\odot$ source of a given type (MS or SB) and a given $\langle U \rangle$ in a given filter. $S_{\textrm{norm}}^{AGN}$ is the same quantity, computed using the \citet{Mullaney2011} AGN template. Note that the filter shape is taken into account for the calculation of $S_\nu$. $p(r_\textrm{AGN})$ is provided by Eq.~\ref{eq:agn} and $p(\langle U \rangle)$ is a log-normal distribution with a width of 0.2~dex (see Sect.~\ref{sect:ingredients}).\\

We compare the predictions of our model with measurements of differential galaxy counts from 24\,$\mu$m to 1.1\,mm (see Fig.\,\ref{fig:counts}). \textit{Spitzer} and \textit{Herschel} counts are well reproduced, showing the effectiveness of our new approach. Note, however, a 10-20\% ($\sim 2 \sigma$) excess at 24\,$\mu$m between 400\,$\mu$Jy and 2\,mJy, and a $\sim$20\% ($\sim 2 \sigma$) excess at the faint-end at 70 and 160\,$\mu$m ($<1$\,mJy and $<5$\,mJy, respectively). The BLAST and SPIRE counts at 250, 350, and 500~$\mu$m are globally well reproduced. Nevertheless, the model slightly overpredicts the three last points of \citet{Bethermin2012} (in red). As discussed by these authors, this could be related to an under-density in GOODS-N. The contribution of lensed sources broadly agrees with the measurements of \citet{Gonzalez-Nuevo2012} at 350\,$\mu$m. At 1.1\,mm, our model nicely agrees with the combined number counts of \citet{Scott2012}\footnote{The counts showed in Fig.\,\ref{fig:counts} are corrected for the bias found in their simulation.}, except for the faintest point, originating from 1-2 $\sigma$ sources and potentially poorly de-biased.\\




Since the LF evolution and number counts may be degenerate \citep{Bethermin2012}, galaxy counts split per redshift provide a powerful test of the validity of our model (note that S12 demonstrated that bolometric IR LF are reproduced at z$<$2.5). This observable is close to the monochromatic LF, but requires fewer corrections (K-corrections, V$_{\textrm{max}}$) which could bias the results (possible biases from photometric redshifts and source identification are discussed in \citealt{Berta2011} and \citealt{Bethermin2012}). The comparison between our model and observations (Fig.\,\ref{fig:counts_zslice}) reveals a good overall agreement between predictions and data. However, we slightly over-predict the counts around 500~$\mu$Jy between z=0.5 and z=2 at 24~$\mu$m. It could be due to a slight excess of PAH features around 15\,$\mu$m in the SB templates. We also underpredict the counts at 100 and 160\,$\mu$m by 1-2\,$\sigma$, probably due to a slight lack of warm dust in the SED templates. Finally, our model overpredicts by $\sim3\,\sigma$ the $z>2$ counts in the 2-6\,mJy range. As explained in the previous paragraph, this could be due to cosmic variance, as these points rely exclusively on GOODS-N.\\


By distinguishing between MS and SB activity, the 2SFM framework allows us to explore selection biases toward MS or SB objects in surveys probing various wavelengths and flux density regimes. MS galaxies (dotted line in Fig.\,\ref{fig:counts}) dominate the number counts at all flux densities and all wavelengths. However, the relative contribution of SBs varies a lot with flux density and wavelength and is important ($\sim$30\%) around 30~mJy at 70\,$\mu$m and 50~mJy at 350 and 500\,$\mu$m. The relative contribution of SB is very sensitive to the evolution of their SED, which is few constrained. If $\langle U \rangle$ did not evolve with redshift, SBs would dominate around 100\,mJy at 350 and 500\,$\mu$m and at flux densities larger than 8\,mJy at 1.1\,mm.\\


Finally, by assuming a non-evolving IR-radio correlation ($q_{TIR} = \textrm{log} \left ( \frac{LIR}{3.75 \times10^{12}\,\textrm{W}} \times \frac{\textrm{W\,Hz}^{-1}}{L_{1.4\,\textrm{Ghz}}} \right) = 2.64$) out to high redshift \citep[e.g.][]{Sargent2010} and a synchrotron spectral slope $\alpha$=0.8 ($S_\nu \propto \nu^{- \alpha}$), we also investigate the contribution of SF galaxies to radio source counts at 1.4\,GHz (see Fig.~\ref{fig:counts}). We combined our model for star-forming objects with the model of AGN-driven radio sources of \citet{Massardi2010}. The result agrees with the compilation of \citet{Vernstrom2011} (see Fig.\,\ref{fig:counts}). According to our model, the 1.4\,GHz counts are dominated by SF objects below 200~$\mu$Jy, in agreement with the observations of e.g. \citet{Seymour2008}. We predict the presence of a bump in the Euclidian-normalized radio counts around 40\,$\mu$Jy which is essentially due to MS galaxies.\\

\section{Conclusion}

\label{sect:conclusion}

Our model based on the main assumption of two SF modes (MS and SB) is able to accurately reproduce the emission of galaxies integrated over most of the Hubble time as probed by galaxy counts from the mid-IR to radio wavelengths. This model contains two main ingredients: the evolution of MS and SB galaxies based on the S12 formalism and a new library of MS and SB SEDs derived from \textit{Herschel} observations (M12). Despite its simplicity, our model provides one of the best fits achieved so far to the number counts, including counts per redshift slice in the SPIRE bands, which were poorly reproduced by the previous generation of models. All these results were obtained without any arbitrary tuning of parameters that are not constrained by observations, contrary to most previous models. The decomposition into 2 modes of SF (2SFM), i.e. MS and SB, associated with two different families of SEDs, is thus a very powerful framework to statistically describe the dust emission of galaxies across cosmic time. In addition, we present a new stochastic AGN treatment, and also found that MS galaxies are responsible for a bump in the 1.4~GHz radio counts around 50~$\mu$Jy.\\


This model can be combined with halo models assuming a link between SFR, M$_\star$, and halo mass \citep[e.g.][]{Bethermin2012a,Wang2012} to interpret the clustering of infrared galaxies and the fluctuation of the cosmic infrared background \citep[e.g.][]{Planck_CIB}. Finally, this model and its future extensions will provide predictions for the next generation of IR, mm, and radio surveys, and, in particular to anticipate which galaxy populations will be preferentially detected, depending on the survey strategy adopted.\\


\acknowledgments{We acknowledge Kimberley Scott, Herv\'e Aussel, Emeric Le Floc'h, Benjamin Magnelli, the anonymous referee, ERC-StG UPGAL 240039, and ANR-08-JCJC-0008.}




\end{document}